# Network Together: Node Classification via Cross-Network Deep Network Embedding

Xiao Shen, Quanyu Dai, Sitong Mao, Fu-Lai Chung, and Kup-Sze Choi

*Abstract*—Network embedding is a highly effective method to learn low-dimensional node vector representations with original network structures being well preserved. However, existing network embedding algorithms are mostly developed for a single network, which fails to learn generalized feature representations across different networks. In this article, we study a cross-network node classification problem, which aims at leveraging the abundant labeled information from a source network to help classify the unlabeled nodes in a target network. To succeed in such a task, transferable features should be learned for nodes across different networks. To this end, a novel cross-network deep network embedding (CDNE) model is proposed to incorporate domain adaptation into deep network embedding in order to learn label-discriminative and network-invariant node vector representations. On the one hand, CDNE leverages network structures to capture the proximities between nodes within a network, by mapping more strongly connected nodes to have more similar latent vector representations. On the other hand, node attributes and labels are leveraged to capture the proximities between nodes across different networks by making the same labeled nodes across networks have aligned latent vector representations. Extensive experiments have been conducted, demonstrating that the proposed CDNE model significantly outperforms the state-of-the-art network embedding algorithms in cross-network node classification.

*Index Terms*— Cross-network embedding, cross-network node classification, deep learning, deep network embedding, domain adaptation, network transfer learning.

## I. Introduction

DOMAIN adaptation aims to transfer the knowledge pre-learned from a source domain to assist in solving the same task in a target domain [1]. Domain adaptation has been widely applied to computer vision (CV) [2]–[5] and natural language processing (NLP) [6], [7]. However, applying domain adaptation to graph mining, such as classifying nodes across networks, has not been sufficiently investigated. Addressing the cross-network node classification problem can benefit various real-world applications. For example, in social network analysis, given a source network where all users are associated with some labels indicating their interest groups, and a target network where only very few users have observable labels. Then, one may want to leverage the abundant labeled information from the source network to make appropriate group recommendations to unlabeled users in the target network. In addition, in protein–protein interaction networks, given a source network with all proteins having annotated functional labels and a target network short of labels, one can take advantage of the rich labeled information from the source network to help identify the functionality of proteins in the target network. To succeed in such cross-network node classification tasks, it is required to learn transferable features for nodes across different networks.

Network embedding is a highly effective method to learn low-dimensional node vector representations with original network structures and properties being well preserved. Most previous network embedding algorithms simply consider plain network structures [8]–[16], which only capture topological proximities between nodes within a network. Recently, a family of attributed network embedding algorithms [17]–[26] has been proposed to capture proximities between nodes by jointly utilizing network structures, node attributes and node labels (if available). Here, node labels refer to the classification labels, and node attributes represent the input features for node classification. Intuitively, the same labeled nodes from different networks might be more likely to have similar attributes than having similar topological structures, especially when node labels depend more on homophily effect [27], rather than structural identity [28]. For example, the articles belonging to the same research area (i.e., labels), say "information security," from different citation networks, might be likely to include some common keywords in their titles (i.e., attributes), such as "privacy, verification, encryption, decryption, and cryptography," while they might have rather distinct topological structures in different networks. Thus, the attributed network embedding algorithms which can capture both attribute affinity and topological proximity should be more suitable for cross-network node classification, as compared with the algorithms solely based on plain network structures.

However, addressing the cross-network node classification problem still faces the following challenges: 1) how to incorporate the heterogeneous data (e.g., network structures, node attributes, and node labels) in a principled way such

Manuscript received May 6, 2019; revised November 10, 2019 and January 26, 2020; accepted May 13, 2020. This work was supported in part by the Hong Kong Ph.D. Fellowship Scheme under Grant PF14-11856, in part by the PolyU UGC Project under Grant P0030970, and in part by the Innovative Technology Fund under Grant MRP/015/18. *(Corresponding author: Quanyu Dai.)*

Xiao Shen and Kup-Sze Choi are with the Centre for Smart Health, The Hong Kong Polytechnic University, Hong Kong (e-mail: shenxiaocam@163.com; thomasks.choi@polyu.edu.hk).

Quanyu Dai, Sitong Mao, and Fu-Lai Chung are with the Department of Computing, The Hong Kong Polytechnic University, Hong Kong (e-mail: quanyu.dai@connect.polyu.hk; sitong.mao@connect.polyu.hk; cskchung@comp.polyu.edu.hk).

Color versions of one or more of the figures in this article are available online at http://ieeexplore.ieee.org.

Digital Object Identifier 10.1109/TNNLS.2020.2995483

that the proximities between nodes within a network and across networks can be well captured? 2) How to exploit and relate the knowledge from different networks to learn node vector representations as network-invariant as possible in order to reduce the problem of varied data distributions across networks?

To address the challenging cross-network node classification problem, we propose a novel cross-network deep network embedding (CDNE) model. In CDNE, two stacked autoencoders (SAEs), i.e., one SAE for the source network (SAE_s) and the other SAE for the target network (SAE_t), are employed to learn low-dimensional node vector representations for cross-network node classification. On the one hand, network topological structures are leveraged to capture the proximities between nodes within a network. Specifically, SAE_s and SAE_t would be employed to reconstruct the associated network structural proximity matrix of the source network and of the target network, respectively. In addition, pairwise constraints are incorporated into SAE_s and SAE_t to embed more strongly connected nodes within each network closer in the latent embedding space. On the other hand, to capture the proximities between nodes across different networks, cross-network node attributes are leveraged to predict fuzzy labels for unlabeled nodes in the target network. Then, both the observable labels and predicted fuzzy labels are leveraged to align nodes across networks according to the class information. In the CDNE model, SAE_s is first trained based on the source network data only by mapping the source network nodes belonging to the same class closer while those belonging to completely different classes far apart from each other in order to learn label-discriminative node vector representations. When SAE_s has converged, or the maximum training iteration has been reached, SAE_s would be fixed afterward. Then, the latent representations of the source network learned by SAE_s would be employed as part of the inputs to train SAE_t. The goal of SAE_t is to learn network-invariant node vector representations by aligning the target network nodes to have similar representations w.r.t. the source network nodes associated with the same labels. Finally, both the label-discriminative and network-invariant node vector representations can be learned by CDNE, which significantly benefits the cross-network node classification task. The contributions of this article can be summarized as follows.

1) The proposed CDNE model is among the first to incorporate domain adaptation into deep network embedding to address the challenging cross-network node classification task.
2) By jointly considering network structures, node attributes, and node labels, the proximities between nodes within a network and across different networks can be well captured.
3) Label-discriminative and network-invariant node vector representations can be effectively learned by CDNE.
4) Extensive experimental results in the real-world data sets demonstrate that CDNE significantly outperforms the state-of-the-art algorithms in cross-network node classification.

The rest of this article is organized as follows. Section II reviews the network embedding and network transfer learning algorithms. Section III introduces the detailed framework of CDNE. Section IV reports the experimental results. Section V concludes this article.

## II. RELATED WORK

### A. Network Embedding

A family of network embedding algorithms has been proposed to preserve topological proximities between nodes within a network based on plain network structures. For example, DeepWalk [10] and node2vec [14] employ random walk sampling strategy and the skip-gram with negative sampling (SGNS) model [29] to learn node vector representations with the preservation of neighborhood structure. GraRep [8] factorizes the positive pointwise mutual information (PPMI) matrix [30] via singular value decomposition in order to capture high-order proximities between nodes within $K$ steps. In addition, motivated by the recent success of deep neural networks in feature representation learning, several deep network embedding algorithms [9], [11], [13], [16], [31] have been proposed to leverage an SAE to learn low-dimensional node vector representations which can best reconstruct the original network connections. This family of network embedding algorithms defines proximities based on the similarity of neighborhood structure between nodes; however, the nodes across different networks generally do not have direct network connections (i.e., not sharing common neighborhood). Thus, the network embedding algorithms based on plain network structures would fail to learn generalized feature representations for nodes across different networks [32], [33].

Besides plain network structures, nodes in the real-world networks are often associated with rich attributes. Recently, a family of attributed network embedding algorithms has been proposed to preserve both network topological proximity and node attribute affinity. For example, Chang *et al.* [18] proposed a heterogenous network embedding framework to learn node vector representations based on node contents and linkage structures. Yang *et al.* [19] proposed a matrix factorization framework to learn network representations from textual information and network structures. Zhang *et al.* [20] proposed an ANRL algorithm that employs a neighbor enhancement AE and an attribute-aware skip-gram model to learn node vector representations from both network structures and node attributes.

In addition, some attributed network embedding algorithms focus on a semisupervised learning problem, where a few nodes can have accessible labels in the attributed network. Then, network structures, node attributes and the observable node labels can be jointly leveraged to learn more informative network representations. For example, Huang *et al.* [17] developed an LANE algorithm to jointly project node labels, network structures and node attributes into a unified embedding space via eigenvector decomposition (EVD). Yang *et al.* [26] proposed a planetoid model to jointly predict node labels and neighborhood contexts, where two types of neighborhood contexts are sampled based on network structures and observable labels, respectively. To alleviate noisy effects from outliers, Liang et al. [23] proposed an SEANO model to collectively capture topological proximity, attribute affinity, and label similarity between nodes. In addition, Kipf and Welling [25] developed a GCN model, which is a variant of convolutional neural networks, to jointly consider network structures, node attributes and partially observable labels for semisupervised node classification. Existing attributed network embedding algorithms are mostly developed for a single network. While in the real-world applications, different networks generally have varied data distributions, which would pose an obstacle for adapting a model learned from a source domain to a target domain [1], [5]. Thus, the single-network-based attributed network embedding algorithms without addressing domain discrepancy would have limited performance in cross-network node classification.

On the other hand, some cross-network embedding algorithms [33]–[36] have been proposed to address the network alignment problem, which assumes that some common nodes should be simultaneously involved in the two aligned networks. In contrast, in cross-network node classification, it is not required to share any common nodes or have any network connections between the source network and the target network. In addition, the goal of network alignment is to infer a node mapping between two networks, while cross-network node classification aims to predict node labels in the target network by leveraging the abundant labeled information from the source network. Thus, the existing cross-network embedding algorithms developed for network alignment cannot be directly applied to address the cross-network node classification task.

### B. Transfer Learning Across Networks

Network transfer learning studies how to transfer useful knowledge learned from a source network to assist in the prediction task in a target network. For example, Ye et al. [37] proposed a transfer learning approach to predict the signed label of edges in a target network by leveraging the edge labeled information from a source network. Tang et al. [38] aim to classify the social relationships in a target network by borrowing the knowledge learned from a source network. Shen et al. [39], [40] developed a CNL model to predict seed nodes and inactive edges for influence maximization in a target network, by leveraging the knowledge prelearned from a source network. Fang et al. [41] developed a network transfer learning (NetTr) algorithm for cross-network node classification, which utilizes the nonnegative matrix trifactorization (NMTF) technique to project the label propagation matrices of the source network and the target network into a common latent space.

A few existing algorithms, e.g., CNL [40], NetTr [41], and GraphSAGE [42], can be applied to address the cross-network node classification task. CNL [40] manually selects a common set of explicit topological features for different networks. Instead of using explicit features, NetTr, GraphSAGE, and the proposed CDNE model aim to learn latent representations for cross-network node classification. In NetTr [41], the latent structural features are learned by a matrix factorization approach, while both GraphSAGE [42] and CDNE employ deep neural networks to learn low-dimensional node vector representations based on network structures, node attributes, and node labels. However, GraphSAGE does not consider the domain discrepancy across different networks, while the proposed CDNE model incorporates the maximum mean discrepancy (MMD) constraints into deep network embedding in order to make the learned node vector representations as much network-invariant as possible.

TABLE I
FREQUENTLY USED NOTATIONS AND DESCRIPTIONS

| Notations | Descriptions |
| --- | --- |
| $\mathcal{G}^s, \mathcal{G}^t$ | Source network and target network |
| $n^s, n^t$ | Number of nodes in $\mathcal{G}^s$ and $\mathcal{G}^t$ |
| $X^s, X^t$ | PPMI matrices of $\mathcal{G}^s$ and $\mathcal{G}^t$ |
| $A^s, A^t$ | Node attribute matrices of $\mathcal{G}^s$ and $\mathcal{G}^t$ |
| $Y^s, Y^t$ | Observable node label matrices of $\mathcal{G}^s$ and $\mathcal{G}^t$ |
| $\hat{Y}^t$ | Predicted node label matrix of $\mathcal{G}^t$ |
| $\mathbb{C}$ | Number of label categories in $\mathcal{G}^s$ and $\mathcal{G}^t$ |
| $L$ | Number of layers of SAE in SAE_s and SAE_t |
| $d(l)$ | Hidden dimensionality of $l$-th layer of SAE_s and SAE_t |
| $v_i^s, v_j^t$ | $i$-th node in $\mathcal{G}^s$ and $j$-th node in $\mathcal{G}^t$ |
| $H^{s(l)}$ | Source network representation learned by $l$-th layer of SAE_s |
| $H^{t(l)}$ | Target network representation learned by $l$-th layer of SAE_t |
| $H_i^{s(l)}$ | Node representation of $v_i^s$ learned by $l$-th layer of SAE_s |
| $H_j^{t(l)}$ | Node representation of $v_j^t$ learned by $l$-th layer of SAE_t |

## III. CROSS-NETWORK DEEP NETWORK EMBEDDING

In this section, we first formulate the cross-network node classification problem and then elaborate on the framework of the proposed CDNE model. For clarity, the frequently used notations are summarized in Table I.

### A. Problem Statement

Let $\mathcal{G}^s = (V^s, E^s, Y^s)$ be a fully labeled source network, with a set of all labeled nodes $V^s$ and a set of edges $E^s$. $Y^s \in R^{n^s \times \mathbb{C}}$ is a label matrix associated with $G^s$, where $n^s = |V^s|$ is the number of nodes in $G^s$ and $\mathbb{C}$ is the number of node categories. $Y_{ic}^s = 1$ if node $v_i^s \in V^s$ is associated with label $c$; otherwise, $Y_{ic}^s = 0$. A node can have multiple labels.

Let $\mathcal{G}^t = (V^t, E^t, Y^t)$ be an insufficiently labeled target network with a set of nodes $V^t = \{V_L^t, V_U^t\}$ and a set of edges $E^t$, where $n^t = |V^t|$ denotes the number of nodes in $\mathcal{G}^t$, $V_L^t$ indicates a very small set of labeled nodes and $V_U^t$ represents a much larger set of unlabeled nodes in $\mathcal{G}^t$. $Y^t \in R^{n^t \times \mathbb{C}}$ is the observable label matrix associated with $G^t$, where $Y_{ic}^t = 1$ if node $v_i^t \in V^t$ has an observable label $c$; otherwise, $Y_{ic}^t = 0$. As the target network has very scarce labeled nodes, the observable label matrix $Y^t$ would be rather sparse.

In addition, nodes in a network can be associated with some attributes. Let $\mathcal{A}^s$ and $\mathcal{A}^t$ denote the sets of node attributes in $\mathcal{G}^s$ and $\mathcal{G}^t$, respectively, where $\mathbb{W}^s = |\mathcal{A}^s|$ and $\mathbb{W}^t = |\mathcal{A}^t|$ represent the number of node attributes in $\mathcal{G}^s$ and $\mathcal{G}^t$. The nodes from the source network and the target network might

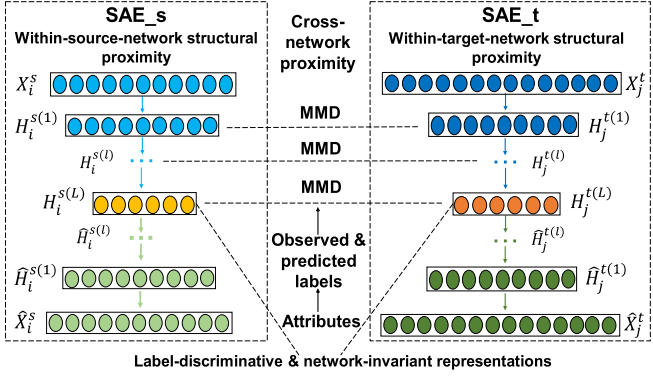

Fig. 1. Framework of the CDNE model. SAE_s and SAE_t have the same number of layers and the same dimensionality for each $l$th ($\forall 1 \leq l \leq L$) hidden layer, but have different input dimensionalities. $X_i^s \in R^{1 \times n^s}$ and $X_j^t \in R^{1 \times n^t}$ denote the input structural proximity vectors associated with $v_i^s$ in $\mathcal{G}^s$ and $v_j^t$ in $\mathcal{G}^t$, respectively. $H_i^{s(l)}$ and $H_j^{t(l)} \in R^{1 \times d(l)}$ represent the node vector representations of $v_i^s$ and $v_j^t$, learned by the $l$th layer of SAE_s and the $l$th layer of SAE_t, respectively.

not share the same set of attributes, i.e., $\mathcal{A}^s \neq \mathcal{A}^t$, but we can build a union set between $\mathcal{A}^s$ and $\mathcal{A}^t$ as $\mathcal{A} = \mathcal{A}^s \cup \mathcal{A}^t$, where $\mathbb{W} = |\mathcal{A}|$ represents the number of union attributes between $\mathcal{G}^s$ and $\mathcal{G}^t$. Then, we construct two matrices $A^s \in R^{n^s \times \mathbb{W}}$ and $A^t \in R^{n^t \times \mathbb{W}}$ to represent the node attributed values associated with $\mathcal{G}^s$ and $\mathcal{G}^t$, respectively. $A_{iw}^r > 0, r \in \{s, t\}$ represents the value of the $w$th attribute (in the union attribute set $\mathcal{A}$) associated with the $i$th node in $\mathcal{G}^r$, while $A_{iw}^r = 0$ indicates that the $i$th node in $\mathcal{G}^r$ is not associated with the $w$th attribute (in the union attribute set $\mathcal{A}$).

Note that in our defined cross-network node classification problem, the network dimensionality (i.e., number of nodes), and the distributions of network connections and node attributes can be varied across networks. However, the two networks should share the same set of node labels. The goal of cross-network embedding is to learn label-discriminative and network-invariant node vector representations such that the abundant labeled information from $\mathcal{G}^s$ can be successfully leveraged to help classify the unlabeled nodes in $\mathcal{G}^t$.

### B. Framework of CDNE

As shown in Fig. 1, on the one hand, CDNE leverages the topological network structures to capture the proximities between nodes within a network. On the other hand, CDNE utilizes the observable labels, and the pseudo fuzzy labels predicted based on node attributes to capture the proximities between nodes across different networks. CDNE consists of two SAEs. First, SAE_s is employed to learn label-discriminative node vector representations for the source network. Next, by minimizing the MMD between the latent representations learned by each corresponding $l$th layer of SAE_s and SAE_t, the learned node vector representations would be as much network-invariant as possible.

The high-order proximities, which can capture global network structural information, have been shown to be beneficial for learning informative feature representations for graph mining [8], [11]. For word embedding in NLP, the PPMI metric [30] has been widely utilized to measure word similarity based on the co-occurrences of two words in a document. By regarding a node in a network as a word in a document, several state-of-the-art network embedding algorithms [8], [9], [11], [43], [44] have adopted PPMI to measure the topological proximities between nodes within $K$ steps in a network. In addition, the SGNS model [29] utilized by the random walk-based network embedding algorithms [10], [14], [19], [21], [45] has been proven to be equivalent to performing factorization on the PPMI matrix [8], [19], [30]. Motivated by this, we employ the PPMI metric to measure the structural proximities between nodes within $K$ steps in a network. Given a network $\mathcal{G}$, its $k$-step transition probability matrix $\mathcal{T}^{(k)} \in R^{n \times n}$ can be obtained via [9], [45], where $n$ is the number of nodes in $\mathcal{G}$, and $\mathcal{T}_{ij}^{(k)}$ indicates the transition probability of visiting node $v_j$ from node $v_i$ after exactly $k$ steps in $\mathcal{G}$. Then, based on a series of $k$-step transition probability matrices up to the maximum $K$-step, i.e., $\{\mathcal{T}^{r(k)}\}_{k=1}^{K}$, we can aggregate an overall transition probability matrix by weighting closer neighborhood more [11]: $\mathcal{T} = \sum_{k=1}^{K} \mathcal{T}^{(k)}/k$. Next, the PPMI [30] metric is employed to measure structural proximities between a pair of nodes as: $X_{ij} = \max(\log[\bar{\mathcal{T}}_{ij}/(\sum_{g=1}^{n} \bar{\mathcal{T}}_{gj}/n)], 0)$, where $\bar{\mathcal{T}}$ is the row-wised normalized transition probability matrix. $X \in R^{n \times n}$ is the PPMI matrix associated with network $\mathcal{G}$, where $X_{ij} > 0$ iff $v_i$ has a strong network connection toward $v_j$ within $K$ steps in $\mathcal{G}$; otherwise, $X_{ij} = 0$.

### C. SAE_s in Source Network

*1) Preserving Source-Network Structural Proximities:* An SAE consists of multiple layers of basic AEs by wiring the hidden representations learned by each layer of AE to the inputs of the successive layer of AE. Given the PPMI matrix of the source network, i.e., $X^s \in R^{n^s \times n^s}$ as the input, an $L$-layer SAE_s is constructed as follows:

$$H^{s(l)} = f\left(H^{s(l-1)}\left(W_1^{s(l)}\right)^T + B_1^{s(l)}\right), \quad l = 1, \ldots, L \quad (1)$$
$$\hat{H}^{s(l-1)} = f\left(\hat{H}^{s(l)}\left(W_2^{s(l)}\right)^T + B_2^{s(l)}\right), \quad l = L, \ldots, 1 \quad (2)$$

where (1) and (2) represent the encoding and decoding process of SAE_s, respectively. $H^{s(0)} = X^s$ is the input PPMI matrix of SAE_s. $H^{s(l)} \in R^{n^s \times d(l)}, \forall 1 \leq l \leq L$ denotes the latent network matrix representation learned by the $l$th layer of SAE_s, and $d(l)$ is the hidden dimensionality of the $l$th layer of SAE_s. The $i$th row of $H^{s(l)}$, denoted as $H_i^{s(l)} \in R^{1 \times d(l)}$, represents the latent node vector representation of $v_i^s$. $\hat{H}^{s(l)}$ is the reconstructed matrix of $H^{s(l)}$ and $\hat{H}^{s(L)} = H^{s(L)}$. In addition, $W_1^{s(l)} \in R^{d(l) \times d(l-1)}$, $B_1^{s(l)} \in R^{n^s \times d(l)}$, $W_2^{s(l)} \in R^{d(l-1) \times d(l)}$, and $B_2^{s(l)} \in R^{n^s \times d(l-1)}$ refer to the encoding weight, encoding bias, decoding weight, and decoding bias matrices associated with the $l$th layer of SAE_s, respectively. $f$ is a nonlinear activation function and the sigmoid activation function $f(x) = 1/(1 + e^{-x})$ is employed in this article.

By minimizing the reconstruction errors of SAE_s, nodes with a more similar neighborhood structure in the source network would have more similar latent vector representations. In addition, to address the network sparsity issue, we follow [13] to incorporate a penalty matrix $P^{s(l)}$ into the reconstruction errors as:

$$\mathcal{R}^{s(l)} = \frac{1}{2n^s}\|P^{s(l)} \odot (\hat{H}^{s(l-1)} - H^{s(l-1)})\|_F^2 \quad (3)$$

where if $H_{ij}^{s(l-1)} > 0$, $P_{ij}^{s(l)} = \beta > 1$; and if $H_{ij}^{s(l-1)} = 0$, $P_{ij}^{s(l)} = 1$. For the first layer of SAE_s, $\beta$ specifies the ratio of penalty on the reconstruction errors of nonzero input elements (i.e., strong network connections) over that of zero input elements (i.e., weak or unobservable network connections). While for the deeper $l$th ($l > 1$) layers of SAE_s, where all the input elements are positive, we regard $\beta^2$ as the weight to rescale the reconstruction errors in the overall loss function.

In addition, we design the pairwise constraint to make more strongly connected nodes (i.e., with higher network structural proximities) have more similar latent node vector representations, as

$$\mathcal{C}^{s(l)} = \frac{1}{2n^s}\sum_{i=1}^{n^s}\sum_{j=1}^{n^s} X_{ij}^s \|H_i^{s(l)} - H_j^{s(l)}\|^2. \quad (4)$$

By minimizing (3) and (4), the network structural proximities between nodes within the source network can be well preserved by the latent node vector representations.

*2) Label-Discriminative Representations:* Next, a matrix $O^s \in R^{n^s \times n^s}$ is defined to represent whether two nodes in $G^s$ share common labels or not. Specifically, $O_{ij}^s = -1$ if $v_i^s$ and $v_j^s$ do not share any common labels, and $O_{ij}^s \geq 1$ indicates the number of common labels shared by $v_i^s$ and $v_j^s$. Then, the following pairwise constraint is devised to learn label-discriminative node vector representations:

$$\mathcal{L}^{s(l)} = \frac{1}{2n^s}\sum_{i=1}^{n^s}\sum_{j=1}^{n^s} O_{ij}^s \|H_i^{s(l)} - H_j^{s(l)}\|^2. \quad (5)$$

Minimizing (5) makes nodes sharing more common labels have more similar latent vector representations while making nodes belonging to completely different categories have rather different latent vector representations.

By integrating the reconstruction errors (3), the pairwise constraint on strongly connected nodes (4), the pairwise constraint on labeled nodes (5), and an *L2*-norm regularization to prevent overfitting $\Omega^{s(l)} = (\|W_1^{s(l)}\|_F^2 + \|W_2^{s(l)}\|_F^2)/2$, the overall loss function of the $l$th layer of SAE_s is defined as

$$\mathcal{J}^{s(l)} = \mathcal{R}^{s(l)} + \alpha^{s(l)}\mathcal{C}^{s(l)} + \varphi^{s(l)}\mathcal{L}^{s(l)} + \lambda^{s(l)}\Omega^{s(l)} \quad (6)$$

where $\alpha^{s(l)}$, $\varphi^{s(l)}$, and $\lambda^{s(l)}$ are the tradeoff parameters to balance the effects of different terms in SAE_s.

### D. SAE_t in Target Network

*1) Preserving Target-Network Structural Proximities:* Similar to SAE_s, an $L$-layer SAE is constructed for SAE_t. Note that the hidden dimensionality of each $l$th layer of SAE_t, i.e., $d(l), \forall 1 \leq l \leq L$, are set the same as in SAE_s. The input matrix of SAE_t is the PPMI matrix of the target network, i.e., $H^{t(0)} = X^t \in R^{n^t \times n^t}$. Note that the input dimensionality of SAE_t is different from SAE_s, i.e., $n^t \neq n^s$. Similar to SAE_s, we devise the reconstruction errors (7) and pairwise constraint (8) to preserve the structural proximities between nodes within the target network, as follows:

$$\mathcal{R}^{t(l)} = \frac{1}{2n^t}\|P^{t(l)} \odot (\hat{H}^{t(l-1)} - H^{t(l-1)})\|_F^2 \quad (7)$$

$$\mathcal{C}^{t(l)} = \frac{1}{2n^t}\sum_{i=1}^{n^t}\sum_{j=1}^{n^t} X_{ij}^t \|H_i^{t(l)} - H_j^{t(l)}\|^2. \quad (8)$$

*2) Network-Invariant Representations:* Next, to learn network-invariant representations, we need to match the distributions of the node vector representations learned for the target network with that of the source network. In domain adaptation, MMD [46] is a widely adopted nonparametric metric to measure the divergence of the distributions between two domains. It has been shown that minimizing MMD in feature representation learning can effectively yield domain-invariant representations [46]–[50]. Motivated by this, we incorporate MMD in SAE_t to learn network-invariant node vector representations. First, the empirical marginal MMD [50] between the source network and the target network is defined as

$$\mathcal{M}_M^{t(l)} = \frac{1}{2}\left\|\frac{1}{n^s}\mathbf{1}^s H^{s(l)} - \frac{1}{n^t}\mathbf{1}^t H^{t(l)}\right\|^2 \quad (9)$$

where $\mathbf{1}^s \in R^{1 \times n^s}$ and $\mathbf{1}^t \in R^{1 \times n^t}$ denote two 1's vectors. By minimizing (9), the marginal distribution of node vector representations learned for the target network can be matched with that of the source network.

Second, the class-conditional MMD [50] between the source network and the target network is defined as

$$\mathcal{M}_C^{t(l)} = \sum_{c=1}^{\mathcal{C}} \frac{1}{2}\left\|\frac{\sum_{i=1}^{n^t} Y_{ic}^t H_i^{t(l)}}{\sum_{i=1}^{n^t} Y_{ic}^t} - \frac{\sum_{j=1}^{n^s} Y_{jc}^s H_j^{s(l)}}{\sum_{j=1}^{n^s} Y_{jc}^s}\right\|^2 \quad (10)$$

where the first term and the second term in (10) represent the average feature vector representation of nodes associated with label $c$ in $G^t$ and $G^s$, respectively, learned by the $l$th layer of SAE_t and the $l$th layer of SAE_s. Note that unlike the source network nodes having completely observable labels, the target network nodes just have very scarce observable labels. Thus, directly utilizing the sparse observable label matrix $Y^t$ in (10) would fail to obtain sufficient statistics to measure the class-conditional distributions of the target network, and then fail to map the same labeled nodes across networks to have similar latent vector representations.

It has been shown that the topological network structures are not generalized across different networks [32], [33]. Intuitively, node attributes are more network-invariant as compared to topological structures, i.e., the same labeled nodes across networks are more likely to have similar attributes than having similar topological structures. Thus, it would be beneficial to leverage the less network-specific attributes to capture cross-network proximities. Specifically, in this article, we propose to utilize cross-network node attributes to predict pseudo labels for unlabeled nodes in the target network. Then, both the observable and pseudo node labels would act as a mean to capture cross-network proximities, by aligning the node vector representations learned by SAE_t to those of SAE_s, according to the class information. Obviously, more accurate pseudo label prediction yields better cross-network alignment. However, the original high-dimensional node attributes might contain noises that would degrade the performance of pseudo label prediction. To address this, we employ principal components analysis (PCA) [51] as a preprocessing step to extract the low-dimensional (i.e., 128-D in the experiments) attribute vector representations. Then, we train a one-versus-rest logistic regression (LR) classifier based on the low-dimensional attribute vector representations of all the labeled nodes from the source network and the target network. Next, the classifier is employed to predict pseudo labels for unlabeled nodes in the target network.

Instead of using pseudo binary labels, we propose to utilize the pseudo fuzzy labels to represent the degree of membership of each node belonging to a specific class. Let $\hat{Y}^t$ denote the predicted label matrix of the target network, where if $v_i^t \in V_L^t$, $\hat{Y}_{ic}^t = Y_{ic}^t \in \{0, 1\}$; and if $v_i^t \in V_U^t$, $0 < \hat{Y}_{ic}^t < 1$ represents the predicted probability of $v_i^t$ to be labeled with category $c$. Next, we replace $Y^t$ with $\hat{Y}^t$ in the conditional MMD loss (10), as in the following:

$$\mathcal{M}_C^{t(l)} = \sum_{c=1}^{\mathcal{C}} \frac{1}{2} \left\| \frac{\sum_{i=1}^{n^t} \hat{Y}_{ic}^t H_i^{t(l)}}{\sum_{i=1}^{n^t} \hat{Y}_{ic}^t} - \frac{\sum_{j=1}^{n^s} Y_{jc}^s H_j^{s(l)}}{\sum_{j=1}^{n^s} Y_{jc}^s} \right\|^2. \quad (11)$$

Note that in (11), a smaller value of $\hat{Y}_{ic}^t$ which indicates $v_i^t$ is predicted as less likely to be labeled with category $c$, would make $v_i^t$ contribute less to the counting of the average latent vector representation of category $c$ for the target network. Thus, by utilizing fuzzy labels to capture different degrees of prediction confidence, the negative effect caused by less confident prediction can be lowered when aligning the cross-network embeddings. By minimizing (11), both the observable and pseudo labeled target network nodes would be aligned to the source network nodes associated with the same labels. In addition, it should be noted that minimizing (5) in SAE_s has already pulled the nodes belonging to different classes far apart from each other. Hence, minimizing (11) in SAE_t would simultaneously make different categories of target network nodes have rather different latent vector representations. Thus, label-discriminative and network-invariant node vector representations can be simultaneously obtained by CDNE.

By integrating the reconstruction errors (7), the pairwise constraints on strongly connected nodes (8), the marginal MMD (9), the class-conditional MMD (11), and an L2-norm regularization $\Omega^{t(l)} = (\|W_1^{t(l)}\|_F^2 + \|W_2^{t(l)}\|_F^2)/2$, the overall loss function of the $l$th layer of SAE_t is defined as

$$\mathcal{J}^{t(l)} = \mathcal{R}^{t(l)} + \alpha^{t(l)} \mathcal{C}^{t(l)} + \mu^{t(l)} \mathcal{M}_M^{t(l)} + \gamma^{t(l)} \mathcal{M}_C^{t(l)} + \lambda^{t(l)} \Omega^{t(l)} \quad (12)$$

where $\alpha^{t(l)}$, $\mu^{t(l)}$, $\gamma^{t(l)}$, and $\lambda^{t(l)}$ are the tradeoff parameters to balance the effects of different terms in SAE_t.

### E. Optimization of CDNE

The whole CDNE model can be trained in three steps, as shown in Algorithm 1. In the first step, SAE_s is layer-wised optimized using stochastic gradient descent (SGD), as in [9] and [11]. When SAE_s has converged, or the

---

**Algorithm 1** CDNE

**Input**: Source network $\mathcal{G}^s = (X^s, Y^s, A^s)$ and target network $\mathcal{G}^t = (X^t, Y^t, A^t)$.

1. Greedy layer-wised training for SAE_s:
  Set $H^{s(0)} = X^s$
  For $l = 1: L$
   1.1 Leverage $H^{s(l-1)}$ as input to $l$-th layer of SAE_s;
   1.2 Given $H^{s(l-1)}, X^s, Y^s$, optimize $l$-th layer of SAE_s by finding $\theta^{s(l)*} = \{W_1^{s(l)*}, W_2^{s(l)*}, B_1^{s(l)*}, B_2^{s(l)*}\} = \arg \mathcal{J}^{s(l)}$ via SGD;
   1.3 Leverage $\theta^{s(l)*}$ to learn $H^{s(l)}$;
  End for
2. Employ PCA on $A^s$ and $A^t$ to extract low-dimensional attribute vector representations and then predict $\hat{Y}^t$;
3. Greedy layer-wised training for SAE_t:
  Set $H^{t(0)} = X^t$
  For $l = 1: L$
   3.1 Leverage $H^{t(l-1)}$ as input to $l$-th layer of SAE_t;
   3.2 Given $H^{t(l-1)}, X^t, \hat{Y}^t, H^{s(l-1)}, Y^s$, optimize $l$-th layer of SAE_t by finding $\theta^{t(l)*} = \{W_1^{t(l)*}, W_2^{t(l)*}, B_1^{t(l)*}, B_2^{t(l)*}\} = \arg \mathcal{J}^{t(l)}$ via SGD;
   3.3 Leverage $\theta^{t(l)*}$ to learn $H^{t(l)}$;
  End for

**Output**: Label-discriminative and network-invariant node vector representations for $\mathcal{G}^s$ and $\mathcal{G}^t$, i.e., $H^{s(L)}$ and $H^{t(L)}$.

---

maximum training iteration has been reached, SAE_s would be fixed afterward. In the second step, the fuzzy labels for unlabeled nodes in the target network can be predicted based on the low-dimensional attribute vector representations extracted by PCA. In the third step, given the node vector representations learned by SAE_s and the observable and predicted node labels as parts of inputs, SAE_t is layer-wise optimized by SGD. Finally, the deepest latent representations learned by SAE_s and SAE_t would be employed as the cross-network node vector representations of CDNE. The time complexity of training SAE_s and SAE_t is $O(n^s c^s hi + n^t c^t hi)$, where $c^s \ll n^s$ and $c^t \ll n^t$ represent the average number of *strongly* connected neighbors (within $K$ steps) per node in the source network and in the target network, respectively. $h = d(1)$ represents the maximum hidden dimensionality in SAE_s and SAE_t, and $i$ refers to the number of training iterations. The time complexity of PCA is $O(\mathbb{W}^2(n^s + n^t) + \mathbb{W}^3)$, where $\mathbb{W}$ indicates the number of union attributes between the source network and the target network. The overall time complexity of CDNE is $O(n^s c^s hi + n^t c^t hi + \mathbb{W}^2(n^s + n^t) + \mathbb{W}^3)$. Since $c^s hi, c^t hi$, and $\mathbb{W}^2$ are independent of $n^s + n^t$, the time complexity of CDNE is linear to the total number of nodes in the source network and the target network.

### F. Interpreting Embeddings Learned by CDNE

On the one hand, in $\mathcal{G}^s$, minimizing (3) and (4) makes strongly connected nodes within $K$ steps have similar latent vector representations. In addition, minimizing (5) yields label-discriminative latent vector representations for the source network nodes.

TABLE II
STATISTICS OF THE REAL-WORLD NETWORKED DATA SETS

| Dataset | #Nodes | #Edges | #Attributes | #Union Attributes | #Labels | Label Distribution (%) | | | | | |
|---|---|---|---|---|---|---|---|---|---|---|---|
| | | | | | | 1 | 2 | 3 | 4 | 5 | 6 |
| Blog1 | 2300 | 33471 | 8189 | 8189 | 6 | 17.35 | 13.91 | 18.22 | 17.04 | 16.39 | 17.09 |
| Blog2 | 2896 | 53836 | 8189 | | | 16.78 | 15.78 | 17.33 | 16.09 | 17.78 | 16.23 |
| | | | | | | 1 | 2 | 3 | 4 | 5 | |
| Citationv1 | 8935 | 15113 | 5379 | 6775 | 5 | 25.32 | 26.02 | 23.92 | 7.88 | 18.72 | |
| DBLPv7 | 5484 | 8130 | 4412 | | | 21.66 | 32.97 | 23.83 | 6.05 | 15.75 | |
| ACMv9 | 9360 | 15602 | 5571 | | | 20.47 | 30.68 | 27.24 | 8.74 | 19.07 | |

On the other hand, in $\mathcal{G}^t$, for a labeled node $v_i^t \in V_L^t$, minimizing (11) makes $v_i^t$ have the latent vector representation similar to the nodes associated with the same labels in $\mathcal{G}^s$, according to the observable binary labels of $v_i^t$. For an unlabeled node $v_i^t \in V_U^t$, minimizing (11) still enables $v_i^t$ to have the latent vector representation similar to the same categories of nodes in $\mathcal{G}^s$, according to the predicted fuzzy labels of $v_i^t$ based on the cross-network node attributes. In addition, if $v_i^t$ is connected to some labeled nodes in $V_L^t$ within $K$ steps, then minimizing (7) and (8) makes $v_i^t$ have the latent vector representation similar to its labeled neighbors in $\mathcal{G}^t$.

By integrating network structures, node attributes and node labels, CDNE can make: 1) more strongly connected nodes within a network have more similar latent vector representations and 2) nodes associated with same labels (within a network and across networks) have similar latent vector representations while nodes associated with different labels (within a network and across networks) have distinct latent vector representations. With such label-discriminative and network-invariant node vector representations, it is beneficial to leverage the abundant labeled information from the source network to help classify the unlabeled nodes in the target network.

## IV. EXPERIMENTS

In this section, we empirically evaluate the performance of the proposed CDNE model.

### A. Experiment Setup

*1) Data Sets:* Experiments were conducted on five real-world networked data sets, as shown in Table II. Blog1 and Blog2 are two disjoint subnetworks extracted from the BlogCatalog[1] data set [52], where a node represents a blogger, and an edge indicates the friendship between two bloggers. Each node is associated with some attributes, i.e., the keywords extracted from the blogger's self-description. Each node is associated with one label indicating the blogger's interest group. Since the two networks were extracted from the same original network, they share the same set of node attributes, and the attribute distributions between two networks are quite similar. To enlarge cross-network distribution discrepancy, in each network, we randomly altered a proportion $p$ of nonzero attributed values to zero and randomly altered the same proportion $p$ of zero attributed values to be "1" in order to simulate incomplete and noisy attributed information across networks, where $p \in \{10\%, 20\%, 30\%, 40\%, 50\%\}$.

On the other hand, Citationv1, DBLPv7, and ACMv9 are three citation networks extracted from the ArnetMiner[2] data sets [53], with different original sources, i.e., Microsoft Academic Graph, DBLP Computer Science Bibliography (DBLP), and Association for Computer Machinery (ACM), respectively. In these data sets, each node represents an article, and each edge indicates the citation of one article to another. We modeled the citation networks as undirected networks. An article can have multiple labels indicating its relevant research topics. The sparse bag-of-words features extracted from the article title were utilized as the attributes for each article. As the three citation networks were extracted from different original sources and formed in different time periods, they do not share a common set of node attributes, and the attribute distributions across these networks are inherently varied to some extent. In the experiments, we constructed two cross-network node classification tasks between Blog1 and Blog2, and six cross-network node classification tasks among Citationv1, DBLPv7, and ACMv9.

*2) Baselines:* The proposed CDNE model was benchmarked against the following baselines. According to the type of information utilized to learn low-dimensional node vector representations, these baselines can be categorized as four types.

*Plain Network Structures:* **DeepWalk** [10] employs random walk sampling and skip-gram model to learn node vector representations with neighborhood preservation. **DNE-APP** [11] is an extension of SDNE [13], which utilizes a semisupervised SAE to reconstruct the PPMI matrix and map nodes with higher topological proximities closer.

*Attributes:* **PCA** [51] extracts low-dimensional attribute vector representations from the original attribute matrix. **TCA** [54] is a domain adaptation algorithm that employs EVD to project the original feature space into a latent space, where the MMD between the source and target domain data can be minimized.

*Network Structures and Attributes:* **NetTr** [41] employs NMTF to project the label propagation matrices of the source network and the target network into a common latent space in order to learn the shared latent structural features. Then, both the latent structural features and node attributes are employed as features for cross-network node classification. **ANRL** [20] leverages a neighbor enhancement AE and an attribute-aware skip-gram model to learn node vector representations based on network structures and node attributes.

---

[1] https://github.com/xhuang31/LANE

[2] https://www.aminer.cn/citation

*Network Structures and Attributes and Labels:* **LANE** [17] employs EVD to jointly project node labels, network structures and node attributes into a unified embedding space. **SEANO** [23] designs a dual-input and dual-output deep neural network architecture to utilize the attributes of each node and its neighborhoods to jointly predict node labels and contexts. **GCN** [25] utilizes a graph convolutional neural network to predict node labels by jointly considering network structures, node attributes and node labels. **GraphSAGE** [42] is an inductive network representation learning algorithm, and its supervised version jointly utilizes node attributes, neighborhood structures, and node labels to learn node vector representations. In the experiments, we adapted GraphSAGE to the transductive setting to make it have better performance in cross-network node classification.

*3) Implementation Details:* In the proposed CDNE model, we built a 2-layer SAE for both SAE_s and SAE_t, with the hidden dimensionality set as $d(1) = 256$ and $d(2) = 128$ for the 1st and 2nd layers of SAE. In SAE_s and SAE_t, we set the maximum step of neighbors as $K = 3$, the ratio of reconstruction error penalty as $\beta = 4$, the weight of $L2$-norm regularization as $\lambda^{s(l)} = \lambda^{t(l)} = 0.05, \forall l \in \{1, 2\}$, and the weight of pairwise constraints on strongly connected nodes as $\alpha^{s(1)} = \alpha^{t(1)} = \alpha = 4$ for the first layer of SAE_s and SAE_t and $\alpha^{s(l)} = \alpha^{t(l)} = \alpha/2, \forall l > 1$ for the deeper layers of SAE_s and SAE_t. In SAE_s, we set the weight of pairwise constraints on labeled nodes as $\varphi^{s(1)} = \varphi = 2$ and $\varphi^{s(l)} = 0, \forall l > 1$. In SAE_t, we set the weight of marginal MMD as $\mu^{t(1)} = \mu = 2$ and $\mu^{t(l)} = \mu/2, \forall l > 1$, and set the weight of conditional MMD as $\gamma^{t(1)} = \gamma = 40, \gamma^{t(l)} = \gamma/2, \forall l > 1$.

The dimensionality of node vector representations learned by each baseline is set the same as in CDNE, i.e., $d = 128$. For a fair comparison, we utilize the PPMI matrix with the same $K$-step to capture network connections for LANE, DNE-APP, and CDNE. Besides, DeepWalk, ANRL, and SEANO all leverage the skip-gram model [29] to sample the truncated random walks to capture network connections between nodes, which has been proven to be equivalent to performing factorization on the PPMI matrix [8], [19], [30]. In addition, for the baselines originally developed for a single-network scenario, we construct a unified network, where the first $n^s$ nodes are from the source network, the last $n^t$ nodes are from the target network, and all the network connections within the source network and within the target network keep remained in the unified network. Then, by utilizing the unified network as the input, the single-network embedding algorithms can be tailored to cross-network node classification, by capturing cross-network proximities based on attribute affinity and/or label similarity in the unified network.

*4) Evaluation Metrics:* To evaluate the cross-network node classification performance, we adopted Micro-F1 and Macro-F1 [55] as two metrics, which have been widely utilized to evaluate the multilabel node classification performance by the network embedding algorithms [10], [13], [20]. Let TP($c$), FP($c$), and FN($c$) denote the number of true positives, false positives, and false negatives associated with label $c$. Micro-F1 gives equal weight to each instance and is defined as follows:

$$\text{Pr} = \frac{\sum_{c=1}^{\mathbb{C}} \text{TP}(c)}{\sum_{c=1}^{\mathbb{C}} \text{TP}(c) + \text{FP}(c)}$$

$$\text{Re} = \frac{\sum_{c=1}^{\mathbb{C}} \text{TP}(c)}{\sum_{c=1}^{\mathbb{C}} \text{TP}(c) + \text{FN}(c)}$$

$$F1_{\text{Micro}} = \frac{2 * \text{Pr} * \text{Re}}{\text{Pr} + \text{Re}}.$$

On the other hand, Macro-F1 gives equal weight to each class and is defined as [56]

$$\text{Pr}(c) = \frac{\text{TP}(c)}{\text{TP}(c) + \text{FP}(c)}, \quad \text{Re}(c) = \frac{\text{TP}(c)}{\text{TP}(c) + \text{FN}(c)}$$

$$F1_{\text{Macro}} = \frac{1}{\mathbb{C}} \sum_{c=1}^{\mathbb{C}} \frac{2 * \text{Pr}(c) * \text{Re}(c)}{\text{Pr}(c) + \text{Re}(c)}.$$

In our setting of cross-network node classification task, all the nodes in the source network have observable labels, and while in the target network, we randomly sample a very small fraction of nodes to give them accessible labels. Then, the low-dimensional node vector representations learned by each baseline are adopted as the features for cross-network node classification. Next, we train a one-versus-rest LR classifier based on all the observable labeled nodes from the source network and the scarce labeled nodes from the target network, and then employ the trained classifier to predict the labels of the unlabeled nodes in the target network. For a specifically labeled fraction in the target network, we generated five random splits of labeled nodes and unlabeled nodes. Then, all the algorithms were evaluated on the same five random splits. The mean and standard deviation of the Micro-F1 and Macro-F1 scores over five random splits were reported for each comparing algorithm.

### B. Performance Comparison

*1) Single-Network Versus Cross-Network Node Classification:* Tables III and IV report the performance of the algorithms for single-network node classification and cross-network node classification, respectively, where only 1% of nodes are with observable labels in the target network. In contrast to cross-network node classification, single-network node classification only leverages the target network data to learn network representations and only leverages the scarce labeled nodes in the target network to train the classifier.

On the one hand, one can observe that for the baselines based on plain network structures, i.e., DeepWalk and DNE-APP, leveraging the cross-network labeled nodes for training would lead to even much lower F1 scores than only utilizing 1% of labeled nodes in the target network for training. This is because the same labeled nodes across networks can have very distinct topological structures. Thus, the network embedding algorithms based on plain network structures are rather unsuitable for cross-network node classification. On the other hand, for the baselines considering node attributes, i.e., PCA, ANRL, LANE, SEANO, and GCN, leveraging cross-network information always yields much better performance than the corresponding single-network node classification tasks. This reflects that node attributes are more generalized across networks as compared with the network topological structures.

TABLE III
SINGLE-NETWORK NODE CLASSIFICATION WHEN ONLY 1% OF NODES ARE LABELED IN THE TARGET NETWORK
(THE NUMBERS IN PARENTHESES ARE THE STANDARD DEVIATIONS OVER FIVE RANDOM SPLITS)

| $\mathcal{G}^t$ | F1 (%) | DeepWalk | DNE-APP | PCA | ANRL | LANE | SEANO | GCN |
|---|---|---|---|---|---|---|---|---|
| Blog1 | Micro | 34.34 (4.13) | 33.8 (3.48) | 23.38 (2.22) | 36.15 (1.4) | 24.37 (4.91) | 24.86 (2.56) | 33.59 (1.41) |
| | Macro | 29.79 (4.38) | 29.2 (5.04) | 22.53 (2.07) | 31.56 (1.88) | 15.49 (5.9) | 21.56 (4.11) | 31.73 (1.92) |
| Blog2 | Micro | 32.72 (3.15) | 38.22 (5.52) | 25.08 (2.77) | 35.79 (1.25) | 27.94 (3.12) | 28.65 (2.45) | 40.56 (2.25) |
| | Macro | 27.17 (4.71) | 33.16 (5.61) | 23.98 (2.14) | 29.87 (2) | 20.92 (3.42) | 24.91 (2.37) | 38.72 (2.31) |
| Citationv1 | Micro | 67.73 (0.99) | 69.64 (1.47) | 46.63 (1.7) | 53.81 (4.06) | 43.14 (2.91) | 71.24 (1.82) | 66.71 (2.03) |
| | Macro | 61.25 (3.49) | 63.97 (2.98) | 39.75 (1.55) | 43.07 (5.54) | 33.68 (3.28) | 67.89 (2.37) | 57.97 (2.24) |
| DBLPv7 | Micro | 57.87 (1.61) | 58.6 (1.06) | 42.56 (1.37) | 47.61 (3.43) | 39.55 (1.34) | 66.54 (2.31) | 58.02 (2.57) |
| | Macro | 49.96 (1.77) | 51.01 (4.11) | 31.8 (3.28) | 36.6 (4.44) | 23.85 (1.89) | 59.28 (3.12) | 47.96 (5.45) |
| ACMv9 | Micro | 59.83 (1.65) | 62.99 (3.15) | 44.35 (2.55) | 49.89 (1.99) | 41.36 (1.43) | 67.59 (1.46) | 61.48 (1.17) |
| | Macro | 59.03 (1.41) | 63.05 (2.95) | 37.35 (2.96) | 41 (3.44) | 29.79 (3.05) | 66.64 (1.41) | 59.04 (1.64) |

TABLE IV
CROSS-NETWORK NODE CLASSIFICATION WHEN ALL NODES ARE LABELED IN THE SOURCE NETWORK AND ONLY 1% OF NODES ARE LABELED IN
THE TARGET NETWORK (THE NUMBERS IN PARENTHESES ARE THE STANDARD DEVIATIONS OVER FIVE RANDOM SPLITS).
THE HIGHEST F1 SCORES AMONG ALL THE COMPARING METHODS ARE SHOWN IN BOLDFACE

| $\mathcal{G}^s \to \mathcal{G}^t$ | F1 (%) | DeepWalk | DNE-APP | PCA | TCA | NetTr | ANRL | LANE | GraphSAGE | SEANO | GCN | CDNE |
|---|---|---|---|---|---|---|---|---|---|---|---|---|
| Blog1→Blog2 | Micro | 28.56 (1.01) | 34.29 (3.76) | 50.63 (0.14) | 52.06 (0.22) | 50.01 (0.2) | 48.42 (0.49) | 54.85 (2.25) | 48.98 (0.72) | 50.17 (0.33) | 53.54 (2.78) | **67.15** (0.51) |
| | Macro | 25.44 (1.34) | 31.68 (4.84) | 49.87 (0.14) | 51.47 (0.23) | 49.05 (0.19) | 42.25 (0.7) | 53.68 (3.4) | 48.35 (0.86) | 50.23 (0.46) | 49.92 (1.81) | **66.99** (0.46) |
| Blog2→Blog1 | Micro | 24.83 (1.75) | 30.81 (2.19) | 52.43 (0.28) | 53.47 (0.27) | 52.32 (0.26) | 46.02 (0.63) | 54.04 (1.16) | 49.84 (0.72) | 49.95 (0.77) | 52.84 (4.38) | **64.64** (0.54) |
| | Macro | 23.23 (2.33) | 27.72 (1.81) | 51.67 (0.27) | 52.77 (0.26) | 51.43 (0.27) | 45.96 (0.45) | 53.12 (1.58) | 48.98 (0.73) | 49.98 (0.74) | 50.14 (4.39) | **64.37** (0.55) |
| Citationv1→DBLPv7 | Micro | 31.3 (2.39) | 53.01 (1.82) | 59.6 (0.13) | 61.54 (0.15) | 59.84 (0.12) | 66.3 (0.19) | 58.53 (0.26) | 71.54 (0.95) | 70.08 (0.42) | 72.11 (0.59) | **74.56** (0.37) |
| | Macro | 27.24 (2.28) | 49.38 (1.95) | 55.38 (0.15) | 57.03 (0.22) | 55 (0.11) | 63.06 (0.21) | 54.87 (0.26) | 67.68 (1.65) | 67.48 (0.65) | 67.53 (0.88) | **71.79** (0.61) |
| DBLPv7→Citationv1 | Micro | 48.47 (2.14) | 63.59 (2.27) | 60.8 (0.06) | 60.14 (0.07) | 59.27 (0.13) | 67.1 (0.18) | 57.89 (0.4) | 71.17 (0.83) | 72.29 (0.22) | 73.21 (0.74) | **80.36** (0.42) |
| | Macro | 42.97 (2.05) | 58.19 (2.34) | 58.03 (0.12) | 56.45 (0.08) | 55.57 (0.25) | 63.85 (0.18) | 54.58 (0.35) | 63.61 (1) | 70.54 (0.4) | 68.92 (1.02) | **78.61** (0.44) |
| Citationv1→ACMv9 | Micro | 39.74 (2.5) | 52.39 (2.8) | 59.15 (0.08) | 59.68 (0.04) | 57.69 (0.06) | 64.65 (0.1) | 56.43 (0.1) | 69.12 (0.49) | 68.4 (0.65) | 71.67 (0.53) | **79.03** (0.54) |
| | Macro | 37.21 (2.91) | 52.24 (2.32) | 55.32 (0.11) | 55.29 (0.05) | 53.37 (0.05) | 62.25 (0.15) | 53.33 (0.25) | 67.26 (0.8) | 67.12 (0.7) | 70.04 (0.83) | **78.47** (0.65) |
| ACMv9→Citationv1 | Micro | 42.6 (1.18) | 54.89 (2.56) | 60.71 (0.16) | 61.89 (0.1) | 58.86 (0.05) | 68.42 (0.15) | 58.53 (0.22) | 73.03 (0.96) | 72.37 (0.43) | 74.14 (0.63) | **79.94** (0.47) |
| | Macro | 36.35 (1.52) | 47.16 (1.34) | 57.87 (0.17) | 58.58 (0.14) | 55.51 (0.06) | 65.61 (0.17) | 55.65 (0.21) | 69.16 (2.13) | 70.6 (0.52) | 70.89 (0.93) | **77.9** (0.52) |
| DBLPv7→ACMv9 | Micro | 38.8 (1.54) | 52.23 (2.11) | 57.07 (0.14) | 57.34 (0.17) | 56.5 (0.2) | 63.41 (0.24) | 54.42 (0.32) | 65.21 (1.16) | 67.45 (0.59) | 69.09 (0.89) | **77.39** (0.58) |
| | Macro | 34.07 (1.98) | 51.18 (2.01) | 52.69 (0.24) | 51.35 (0.19) | 51.42 (0.3) | 60.52 (0.4) | 50.21 (0.4) | 60.74 (2.63) | 66.41 (0.79) | 67.03 (0.78) | **77** (0.7) |
| ACMv9→DBLPv7 | Micro | 40.78 (0.9) | 41.94 (2.82) | 58.37 (0.1) | 59.78 (0.12) | 56.41 (0.16) | 64.52 (0.11) | 57.29 (0.62) | 70.94 (0.42) | 66.16 (0.36) | 68.03 (1.12) | **73.16** (0.34) |
| | Macro | 35.02 (1.46) | 32.58 (2.33) | 53.52 (0.13) | 55.2 (0.12) | 49.87 (0.12) | 61.1 (0.24) | 52.81 (0.57) | 65.4 (1.44) | 63.56 (0.37) | 64.22 (1.75) | **70.95** (0.29) |

In the proposed CDNE model, the network-specific topological structures have been employed to capture within-network proximities, and the relatively network-invariant node attributes have been utilized to capture cross-network proximities. Thus, CDNE is indeed suitable for cross-network node classification.

*2) Incorporate Heterogeneous Data for Network Embedding:* Both NetTr and ANRL leverage network structures and node attributes to construct features for node classification. However, ANRL outperformed NetTr in most cross-network node classification tasks, as shown in Table IV. This is because, in NetTr, the common latent structural features are learned independently of node attributes. While ANRL can capture the correlations between node attributes and neighborhood structures during the network representation learning process. In addition, both DNE-APP and CDNE employ SAE to preserve network structural proximities. While the significant outperformance of CDNE over DNE-APP demonstrates that besides plain network structures, also utilizing node attributes and node labels can effectively improve the network embedding quality. Moreover, the significant improvement of CDNE over PCA reflects that besides node attributes, jointly utilizing network structures and node labels can yield more informative node vector representations. Thus, network structures, node attributes and node labels should be all beneficial for learning informative feature representations for cross-network node classification. However, how to appropriately incorporate such heterogeneous data for cross-network embedding is nontrivial.

*3) Domain Discrepancy Across Networks:* As shown in Table IV, TCA can achieve higher F1 scores than PCA in most tasks. This demonstrates the effectiveness of reducing domain discrepancy on cross-network node classification. However, the performance of TCA is worse than LANE on Blog networks and also much worse than GraphSAGE, SEANO, and GCN on the citation networks. Note that the attributed network embedding algorithms not only utilize node attributes but also take full advantage of network structures when learning node vector representations. However, the conventional domain adaptation algorithms developed for CV or NLP are generally based on the assumption that the data samples in each domain are independent and identically distributed. It has been shown that considering the complex network relationships between nodes should be rather important and necessary for node classification in network structural data [10], [13], [11]. Thus, the conventional domain adaptation algorithms without considering network topological structures would fail to achieve good performance in cross-network node classification.

In addition, as shown in Table IV, CDNE achieves the highest F1 scores among all the comparing algorithms, in all the cross-network node classification tasks. Besides, GCN, GraphSAGE, and SEANO also achieve relatively good performance. Note that GCN, GraphSAGE, SEANO, and CDNE all employ deep neural networks to learn latent node vector representations based on network structures, node attributes and node labels. When only 1% of nodes are with observable labels in the target network, the improvement of CDNE over these baselines is rather significant. This is because although the attributed network embedding algorithms can capture cross-network proximities based on attribute affinity and label similarity, they do not address the varied data distributions across networks, which would pose an obstacle for applying a model trained in the source network to the target network. In contrast, the proposed CDNE model incorporates the MMD constraints into deep network embedding in order to make the learned node vector representations network-invariant. The significant outperformance of CDNE over GCN, GraphSAGE, and SEANO demonstrates that reducing domain discrepancy should be indeed necessary and important for cross-network node classification.

Next, we investigate the performance of the algorithms when the fraction of scarce labeled nodes in the target network is varied in $\{0.5\%, 1\%, 3\%, 5\%, 7\%, 10\%\}$. As shown in Fig. 2(a) and (b), for the tasks between Blog1 and Blog2, when the labeled fraction is 10% in the target network, LANE can almost match the performance of CDNE. In addition, as shown in Figs. 2(c)–(h), for the tasks among the citation networks, the improvement of CDNE over GCN and GraphSAGE becomes less as the labeled fraction of the target network is larger. This is because when more nodes are with observable labels in the target network, more unlabeled nodes would have network connections w.r.t. the labeled nodes and thus have similar latent representations with respect to (w.r.t.) their labeled neighbors. Then, the existing network embedding algorithms which can well capture the proximities between the unlabeled nodes and their labeled neighbors within the target network have already been able to achieve good node classification performance. However, one can see that when the labeled fraction is rather small in the target network, i.e., 0.5% or 1%, the improvement of CDNE over the baselines is very significant. This is because when less labeled nodes are available in the target network, it would be more helpful and necessary to take advantage of the knowledge from the source network. When transferring knowledge across networks, reducing domain discrepancy should be rather important and essential for achieving good performance in the target network. Thus, the proposed CDNE model, which incorporates domain adaptation into deep network embedding, is indeed effective for cross-network node classification, especially when very scarce labeled nodes are available in the target network.

The earlier results of Blog networks (shown in Tables III and IV and Fig. 2) are reported when the proportion of random alternation for node attributes in Blog1 and Blog2 is fixed as $p = 30\%$. Next, we varied the proportion of random alternation $p$ in $\{10\%, 20\%, 30\%, 40\%, 50\%\}$ to investigate how the algorithms respond to different degrees of noise across networks. Note that when a larger proportion of random alternations are added, the cross-network distribution discrepancy of node attributes will be larger. As shown in Fig. 3, larger domain discrepancy always yields worse cross-network node classification performance for all algorithms. However, when the domain discrepancy is larger, the improvement of CDNE over the best baseline is generally more significant. This reflects that the varied data distributions across networks actually hamper the transferal of knowledge across networks. Thus, the proposed CDNE model, which incorporates MMD into deep network embedding, can significantly outperform the network embedding baselines without addressing domain discrepancy.

### C. Ablation Test

To investigate the contributions of different components in the proposed CDNE model for cross-network node classification, we conducted ablation studies and reported the results in Table V. CDNE ($\alpha = 0$) indicates not incorporating the pairwise constraints on strongly connected nodes, i.e., (4) in SAE_s and (8) in SAE_t. CDNE ($\varphi = 0$) represents that the pairwise constraint on labeled nodes (5) is not incorporated into SAE_s. In addition, CDNE ($\mu = 0$) and CDNE ($\gamma = 0$) indicate without incorporating marginal MMD (9) and conditional MMD (11) in SAE_t, respectively.

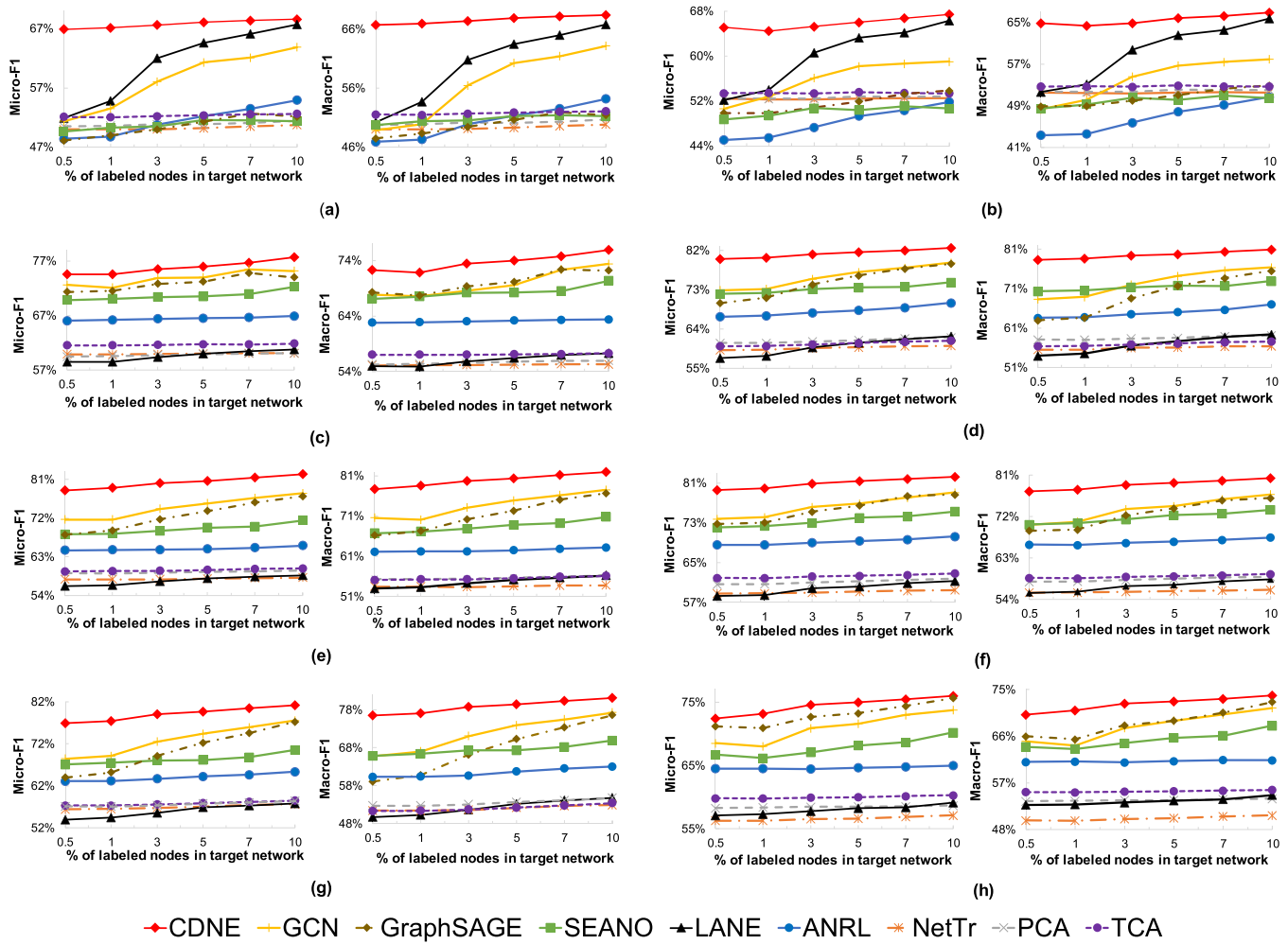

Fig. 2. Cross-network node classification with varied small fractions of labeled nodes in the target network and fully labeled nodes in the source network. (a) From Blog1 to Blog2. (b) From Blog2 to Blog1. (c) From Citationv1 to DBPLPv7. (d) From DBPLPv7 to Citationv1. (e) From Citationv1 to ACMv9. (f) From ACMv9 to Citationv1. (g) From DBPLPv7 to ACMv9. (h) From ACMv9 to DBPLPv7.

TABLE V
MICRO-F1 (%) AND MACRO-F1 (%) OF CDNE VARIANTS FOR CROSS-NETWORK NODE CLASSIFICATION WHEN ALL NODES ARE LABELED IN THE SOURCE NETWORK AND ONLY 1% OF NODES ARE LABELED IN THE TARGET NETWORK

| Model Variants | $\mathcal{G}^s$ / $\mathcal{G}^t$ | Blog1 / Blog2 | Blog2 / Blog1 | Citationv1 / DBLPv7 | DBLPv7 / Citationv1 | Citationv1 / ACMv9 | ACMv9 / Citationv1 | DBLPv7 / ACMv9 | ACMv9 / DBLPv7 |
|---|---|---|---|---|---|---|---|---|---|
| CDNE | Micro | 67.15 | 64.49 | 74.56 | 80.36 | 79.03 | 79.94 | 77.39 | 73.16 |
|  | Macro | 66.99 | 64.37 | 71.79 | 78.61 | 78.47 | 77.90 | 77.00 | 70.95 |
| CDNE ($\alpha = 0$) | Micro | 66.80 | 62.71 | 74.29 | 78.92 | 77.81 | 79.75 | 75.49 | 71.83 |
|  | Macro | 66.20 | 61.97 | 71.65 | 76.95 | 77.06 | 77.66 | 74.99 | 69.38 |
| CDNE ($\varphi = 0$) | Micro | 28.35 | 22.95 | 73.50 | 79.14 | 77.38 | 78.65 | 76.51 | 69.27 |
|  | Macro | 20.23 | 12.08 | 71.31 | 77.12 | 77.21 | 76.48 | 76.04 | 67.00 |
| CDNE ($\mu = 0$) | Micro | 67.38 | 64.49 | 74.62 | 80.32 | 78.91 | 79.73 | 77.65 | 72.65 |
|  | Macro | 67.20 | 64.33 | 71.91 | 78.28 | 78.21 | 77.57 | 77.18 | 70.40 |
| CDNE ($\gamma = 0$) | Micro | 29.46 | 27.33 | 59.47 | 68.55 | 62.25 | 67.71 | 61.08 | 59.99 |
|  | Macro | 26.11 | 23.96 | 52.72 | 61.66 | 61.34 | 61.66 | 60.73 | 51.57 |

As shown in Table V, CDNE ($\alpha = 0$) achieves worse performance than CDNE in five tasks while achieving comparable performance w.r.t. CDNE in the other tasks. This demonstrates that incorporating the pairwise constraints on strongly connected nodes is not always significant for all tasks. This might be because by minimizing the reconstruction errors of PPMI matrices, the within-network topological proximities have already been captured.

Second, CDNE ($\varphi = 0$) always yields much worse performance than CDNE in all tasks. This reflects that learning label-discriminative representations is rather important for cross-network node classification. Moreover, we can see that CDNE ($\gamma = 0$) leads to even worse performance than CDNE ($\varphi = 0$). This demonstrates that learning network-invariant node vector representations by conditional MMD is essential for CDNE to achieve good performance in cross-network node classification. In the conditional MMD, we use the observable binary labels, and the fuzzy labels predicted based on cross-network node attributes to align the same labeled nodes across networks. Without conditional MMD, our model would not take advantage of the useful node attributes, and then the performance would be significantly dropped. Besides, one can observe that CDNE ($\mu = 0$) achieves similar results w.r.t. CDNE, which reflects that the marginal MMD does not show a significant effect on the performance of CDNE. This might be because by minimizing conditional MMD, each category of nodes across networks would have similar latent vector representations, and as a result, the marginal distributions between the source network and the target network in the embedding space can also be minimized.

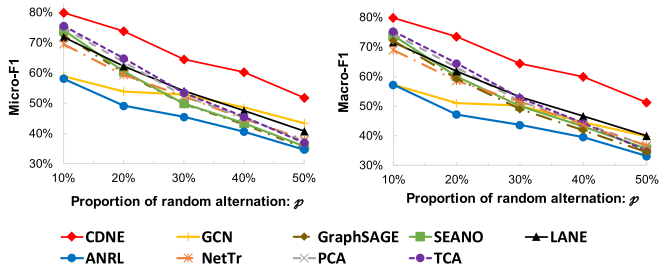

Fig. 3. Cross-network node classification task from Blog2 to Blog1, when varied proportions of random alternation are added to each network to simulate the incomplete and noisy attributed information across networks. All nodes are labeled in the source network, and only 1% of nodes are labeled in the target network.

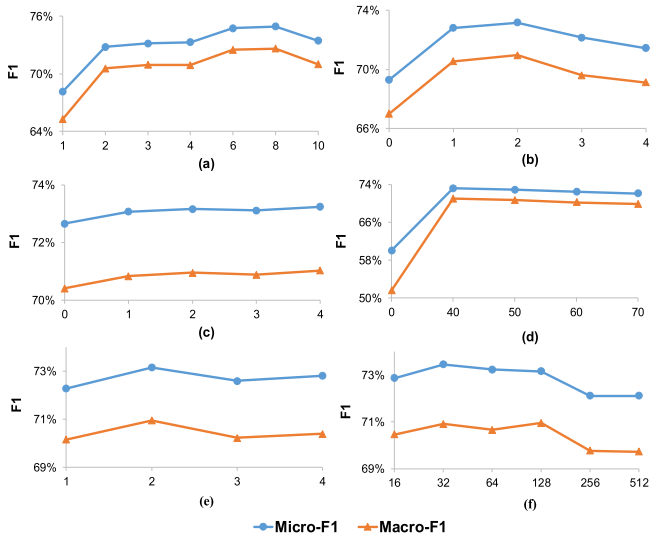

Fig. 4. Sensitivity of the parameters, i.e., $K, \varphi, \mu, \gamma, L$, and $d$ on the performance of CDNE in the cross-network node classification task from ACMv9 to DBLPv7 when all nodes are labeled in the source network, and 1% of nodes are with observable labels in the target network. (a) Value of maximum step: $K$. (b) Weight of pairwise constraint on labeled nodes: $\varphi$. (c) Weight of marginal MMD: $\mu$. (d) Weight of conditional MMD: $\gamma$. (e) #layer of SAE: $L$. (f) #dimension: $d$.

### D. Parameter Sensitivity

Next, we analyze the sensitivities of the parameters, i.e., $K, \varphi, \mu, \gamma, L$, and $d$ on the performance of CDNE.

Parameter $K$ denotes the maximum step of neighbors utilized to measure topological proximities between nodes within a network. As shown in Fig. 4(a), in the citation network, high-order proximities (i.e., a larger value of $K$) can yield better performance. This is because the high-order proximities, which can capture global structural information, are beneficial for learning informative representations for node classification [8], [11]. However, when $K$ is too large (i.e., 10), the performance will decrease, as shown in Fig. 4(a). This is because the neighbors too far away from a target node would become less similar, and mapping such dissimilar nodes to have similar representations would introduce noise for node classification. In the Blog network, as shown in Fig. 5(a), $K = 1$ or $K \geq 3$ can all achieve good performance.

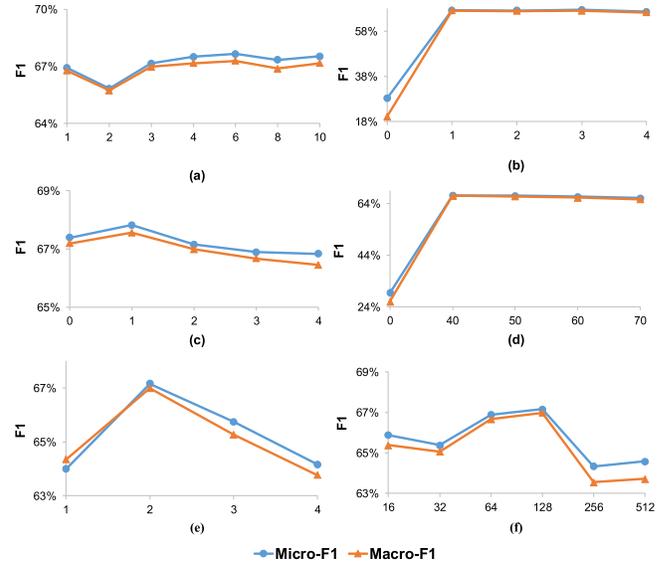

Fig. 5. Sensitivity of the parameters, i.e., $K, \varphi, \mu, \gamma, L$, and $d$ on the performance of CDNE in the cross-network node classification task from Blog1 to Blog2 when all nodes are labeled in the source network, and 1% of nodes are with observable labels in the target network. (a) Value of maximum step: $K$. (b) Weight of pairwise constraint on labeled nodes: $\varphi$. (c) Weight of marginal MMD: $\mu$. (d) Weight of conditional MMD: $\gamma$. (e) #layer of SAE: $L$. (f) #dimension: $d$.

Parameter $\varphi$ denotes the weight of pairwise constraint on labeled nodes in the source network. As shown in Figs. 4(b) and 5(b), $\varphi > 0$ always yields much higher F1 scores than $\varphi = 0$. As shown in Fig. 4(b), in the citation network, $\varphi = 2$ achieves the best performance, but too large values of $\varphi$ would degrade the performance. While in the Blog network, as shown in Fig. 5(b), the performance of CDNE is insensitive to the value of $\varphi$, $\forall \varphi \in \{1, 2, 3, 4\}$.

Parameters $\mu$ and $\gamma$ represent the weight of marginal MMD and conditional MMD in SAE_t, respectively. As shown in Figs. 4(c) and 5(c), the performance of CDNE is not very sensitive to the value of $\mu$. In addition, as shown in Figs. 4(d) and 5(d), $\gamma > 0$ always leads to significantly higher F1 scores than $\gamma = 0$, and the performance of CDNE is insensitive to the value of $\gamma$, $\forall \gamma \in \{40, 50, 60, 70\}$.

Parameter $L$ denotes the number of layers of SAE in SAE_s and SAE_t. As shown in Figs. 4(e) and 5(e), a two-layer SAE can achieve better performance than a shallow architecture, i.e., one-layer basic AE. However, a deeper architecture with more than two layers of SAE would degrade the performance.

Parameter $d$ is the dimensionality of the deepest latent node vector representations learned by SAE_s and SAE_t. As shown in Figs. 4(f) and 5(f), $d \in \{64, 128\}$ leads to good performance for CDNE, while too small or too large dimensionalities would lead to the degraded performance.

## V. Conclusion

In this article, we address a cross-network node classification problem of how to leverage the abundant labeled information from a source network to help classify the unlabeled nodes in a target network, with the source network and the target network characterized by fully labeled nodes and very scarce labeled nodes respectively. A CDNE model is proposed to address the involved challenges. CDNE employs two SAEs to learn the low-dimensional node vector representations for the source network and the target network, respectively. On the one hand, CDNE utilizes network structures, node attributes, and node labels to capture within-network and cross-network proximities. On the other hand, the marginal and class-conditional MMD constraints have been incorporated into CDNE to learn network-invariant node vector representations. Extensive experimental results demonstrate that the proposed CDNE model achieves significant gains over the state-of-the-art network embedding algorithms in cross-network node classification.

For future work, we plan to extend CDNE to address a more generalized task, where the source network and the target network might not share the fully identical node labels. In addition, in this article, the proposed CDNE model focuses on a cross-network node classification task where only one source network and one target network are given. In the future, it is interesting to study the multisource cross-network node classification task where the abundant labeled data from multiple source networks can be leveraged to help classify unlabeled nodes in the target network.

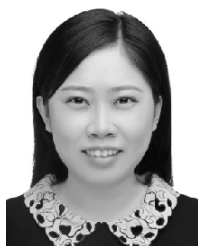

**Xiao Shen** received the B.Sc. degree from the Beijing University of Posts and Telecommunications, Beijing, China, in 2012, the B.Sc. degree from the Queen Mary University of London, London, U.K., in 2012, the M.Phil. degree from the University of Cambridge, Cambridge, U.K., in 2013, and the Ph.D. degree from the Department of Computing, The Hong Kong Polytechnic University, Hong Kong, in 2019.

She is currently a Post-Doctoral Researcher with The Hong Kong Polytechnic University. She has authored or coauthored in prestigious international journals and conferences, including the IEEE TRANSACTIONS ON FUZZY SYSTEMS, IEEE TRANSACTIONS ON CYBERNETICS, IEEE TRANSACTIONS ON NEURAL NETWORKS AND LEARNING SYSTEMS, the International ACM SIGIR Conference on Research and Development in Information Retrieval (SIGIR), the International World Wide Web Conference (WWW), and the AAAI Conference on Artificial Intelligence (AAAI). Her research interests include feature representation learning, deep learning, transfer learning, and data mining in complex networks.

Dr. Shen received the Hong Kong Ph.D. Fellowship.

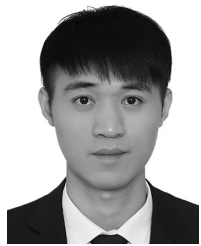

**Quanyu Dai** received the B.Eng. degree from the Department of Electronic Engineering, Shanghai Jiao Tong University, Shanghai, China, in 2015, and the Ph.D. degree from the Department of Computing, The Hong Kong Polytechnic University, Hong Kong, in 2020. He has authored or coauthored in the top-tier conferences, including the International Joint Conferences on Artificial Intelligence (IJCAI), the AAAI Conference on Artificial Intelligence (AAAI), the International World Wide Web Conference (WWW), and the Conference on Information and Knowledge Management (CIKM). His research interests include representation learning, transfer learning, deep learning, and adversarial learning for graph-structured data.

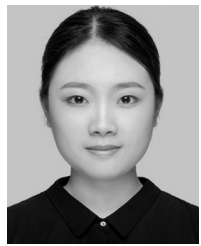

**Sitong Mao** received the B.Eng. degree in computer science from the University of Xiamen, Xiamen, China, in 2015. She is currently pursuing the Ph.D. degree with the Department of Computing, The Hong Kong Polytechnic University, Hong Kong.

Her research interests include deep learning and transfer learning.

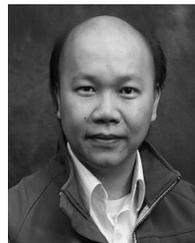

**Fu-Lai Chung** received the B.Sc. degree from the University of Manitoba, Winnipeg, MB, Canada, in 1987, and the M.Phil. and Ph.D. degrees from The Chinese University of Hong Kong, Hong Kong, in 1991 and 1995, respectively.

He is currently an Associate Professor with the Department of Computing, The Hong Kong Polytechnic University, Hong Kong. He has authored or coauthored widely in prestigious international journals, including the IEEE TRANSACTIONS ON NEURAL NETWORKS AND LEARNING SYSTEMS, IEEE TRANSACTIONS ON FUZZY SYSTEMS, IEEE TRANSACTIONS ON CYBERNETICS, IEEE TRANSACTIONS ON KNOWLEDGE AND DATA ENGINEERING, *Pattern Recognition*, and *Neural Networks*. His current research interests include deep learning, transfer learning, adversarial learning, social network analysis and mining, recommendation systems, and big data learning.

Dr. Chung serves on program committees of top international conferences, including the IEEE International Conference on Data Mining, the AAAI Conference on Artificial Intelligence (AAAI), and the International Conference on Pattern Recognition (ICPR).

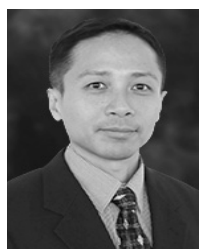

**Kup-Sze Choi** received the Ph.D. degree in computer science and engineering from The Chinese University of Hong Kong, Hong Kong.

He is currently a Professor and the Director of the Centre for Smart Health, The Hong Kong Polytechnic University. His research interests include virtual reality and artificial intelligence, and their applications in medicine and healthcare.